\documentclass[11pt,a4paper]{article}
\usepackage[a4paper,total={6.0in, 8.5in}]{geometry}
\usepackage{graphicx,hyperref,enumitem}
\usepackage[round]{natbib}
\usepackage{amsmath,amssymb,xcolor}

\usepackage[normalem]{ulem}
\title{\textbf{Empirical distributions of the robustified $t$-test statistics}}
\author{
\textbf{Chanseok Park}\thanks{cpark2@gmail.com} \\
Applied Statistics Laboratory\\
Department of Industrial Engineering\\ Pusan National University\\
Busan, Republic of Korea
\and
\textbf{Min Wang}\thanks{min.wang3@utsa.edu}\\
Department of Management Science and Statistics \\
The University of Texas at San Antonio \\
San Antonio, TX, USA
\and
\textbf{Wook-Yeon Hwang}\thanks{wyhwang@dau.ac.kr}   \\
College of Global Business \\
Dong-A University \\
Busan, Republic of Korea
}
\date{}

\begin{document}
%%===============================================
\maketitle
%------------------------------------------------
\begin{abstract}
One sample $t$-test is commonly used for the statistical analysis of experimental data
in comparing the mean difference between matched pairs
such as the pre-treatment and post-treatment assessments. It is well known that
the $t$-test is sensitive to data contamination that occurs frequently in practical applications.
To overcome the non-robustness of the $t$-test, \cite{Park:2018a} developed two robustified analogues of this
test based on robust statistics and provided the asymptotic distributions for the statistics, assuming
that the sample size is large enough. These asymptotic results may not be
adequate for making statistical inference including hypothesis testing
and confidence interval when the sample size is small or even moderate. The purpose of this paper
is to conduct Monte Carlo
simulations to obtain the empirical distributions of these test statistics and their quantiles
to conduct statistical inference. Useful tables are also constructed for a quick finding of
their empirical quantiles with
different sample sizes.

%------------------------------------------------
\bigskip
\noindent\textbf{\uppercase{Keywords:~}}
$t$-test, robustness, median, median absolute deviation, Hodges-Lehmann, Shamos estimator.
\end{abstract}
%------------------------------------------------

%============================
\section{Introduction}
%----------------------------
One sample $t$-test for statistical hypothesis testing is a standard topic in all classes
when teaching undergraduate students basic probability and elementary statistics.
Assuming that the data $X_i$'s ($i =1, \cdots, n$) are independent and identically distributed
(iid) normal random variables with mean $0$ and variance $\sigma^2$. Let
$$
\bar{X} = \frac{\sum_{i =1}^{n}X_i}{n} \quad \mathrm{and} \quad S^2 = \frac{\sum_{i =1}^{n}(X_i - \bar{X})^2}{n-1}
$$
be the sample mean and sample variances of $X_i$'s respectively. The commonly used one sample $t$-test statistic for testing
the null $H_0: \mu = \mu_0$ against the two-sided alternative $H_a: \mu \neq \mu_0$ is defined as
\begin{equation}\label{t:test}
T=  \frac{\bar{X}-\mu_0}{S/\sqrt{n}},
\end{equation}
where $\mu_0$ is a pre-specified target value of the true population mean $\mu$. At the $\alpha$ significance level,
let $t_{1-\alpha/2, \nu}$ be the $1-\alpha/2$ quantile of $T$ distribution
with degrees of freedom $\nu$, denoted by $T_\nu$. We reject the null hypothesis $H_0$ if the absolute value of
the observed statistic in (\ref{t:test}) is larger than $t_{1-\alpha/2, n-1}$ or the resulting two-sided
$p$ value is less than $\alpha$. This test is included in most texts suitable for
the elementary statistics course; see, for example \cite{Weiss:2012}.

We observe that the test statistic in (\ref{t:test}) depends on the sample mean and sample standard deviation, which are very sensitive
to outliers that appear in the experimental data. Thus, the performance of this statistic could deteriorate significantly due to
the lack of robustness of these estimators. An intuitive way to deal with this issue is employ robust estimators, such as
the sample median as an alternative to the sample mean and the median absolute deviation estimator as an alternative for the standard deviation.
On the basis of the above intuition, \cite{Park:2018a} and \cite{Jeong/Son/Lee/Kim:2018} proposed 
two robustified analogues of the $t$-test statistic
based on the robust estimators. Specifically, they developed two robust test statistics
based on the median and the median absolute deviation estimators,
and the Hodges-Lehmann estimator \citep{Hodges/Lehmann:1963}
and the Shamos estimator \citep{Shamos:1976}. It deserves mentioning that the proposed robustified test statistics are
not only simple and easy to implement in practical applications, but also are pivotal quantities and converge
to the standard normal distribution assuming that the sample size is large enough.

However, when the sample sizes are small or even moderate, it may not be
appropriate to use the asymptotic properties of these statistics
for making statistical inference, since the resulting $p$-values and confidence
intervals based on the asymptotic normality may not be accurate, limiting their
practical applications for small sample sizes. To deal with this issue,
we carry out extensive Monte Carlo
simulations to obtain the empirical distributions of these robustified
$t$-test statistics and calculate their related quantiles,  
which can then be used for obtaining accurate statistical inference from the small sample sizes.

The remainder of this paper is organized as follows. In Section \ref{section:02}, we
briefly review the two robustified $t$-test statistics developed by \cite{Park:2018a} and \cite{Jeong/Son/Lee/Kim:2018}.
In Section \ref{section:03}, we carry out extensive Monte Carlo simulations to obtain the
empirical distributions of these two statistics. In Section \ref{section:04},
we provide several examples to illustrate the use of the derived empirical distributions.
Finally, some concluding remarks are provided in \ref{section:05}, with table for the empirical
quantiles deferred to the Appendix.

%============================
\section{Robustified $t$-test statistics} \label{section:02}
%----------------------------
%{For the sake of completeness, in} this section, we briefly  the test statistics
%proposed by \cite{Park:2018a} and \cite{Jeong/Son/Lee/Kim:2018}.

Recently, \cite{Park:2018a} proposed a robustified analog of the $t$-test statistics
in (\ref{t:test}) by replacing its mean and standard deviation
with the median and the median absolute deviation (MAD), respectively, resulting in the
following modified $t$-test statistic
\[
T = \frac{ \hat{\mu}_m - \mu_0}{ \hat{\sigma}_M/\sqrt{n} },
\]
where $\hat{\mu}_m = \mathop{\mathrm{median}}_{1\le i\le n} X_i$ and
$\hat{\sigma}_M = \mathop{\mathrm{median}}_{1\le i\le n}
  \big| X_i - \mathop{\mathrm{median}}_{1\le i\le n} X_i  \big|$.
More importantly, \citeauthor{Park:2018a} showed that the above statistic is a pivotal quantity,
whereas it does not converge to the standard normal distribution. Thus, to achieve the
asymptotic normality of the test statistic, \citeauthor{Park:2018a} suggested the following test statistic
%-------------------------------
\begin{equation} \label{EQ:TA}
T_A =
\sqrt{\frac{2n}{\pi}} {\Phi^{-1}\Big(\frac{3}{4}\Big)} \cdot
\frac{\displaystyle\mathop{\mathrm{median}}_{1\le i\le n}X_i-\mu}%
{\displaystyle\mathop{\mathrm{median}}_{1\le i\le n}
  \big| X_i - \mathop{\mathrm{median}}_{1\le i\le n} X_i  \big|}
\stackrel{d}{\longrightarrow} N(0,1),
\end{equation}
{where $\Phi^{-1}(\cdot)$ is the inverse of the standard normal cumulative distribution function and $\stackrel{d}{\longrightarrow}$ denotes convergence in distribution.}
%-------------------------------

Later on, analogous to the idea of \cite{Park:2018a}, \cite{Jeong/Son/Lee/Kim:2018} proposed
another robustified version of the $t$-test statistic in which
the Hodges-Lehmann estimator \citep{Hodges/Lehmann:1963} and
the Shamos estimator \citep{Shamos:1976} are employed to replace the mean and standard deviation
in (\ref{t:test}, which results in the following test statistic given by
\[
T = \frac{\hat{\mu}_H - \mu_0}{\hat{\sigma}_S/\sqrt{n}},
\]
where {$\hat{\mu}_H$ and $\hat{\sigma}_S$  represent the Hodges-Lehmann and the Shamos estimators, respectively}.
Note that the Hodges-Lehmann estimator is defined as
\[
\hat{\mu}_H
= \mathop{\mathrm{median}}_{i \le j} \Big( \frac{X_i+X_j}{2} \Big)
\]
and the Shamos estimator is defined as
\[
\hat{\sigma}_S
= \displaystyle\mathop{\mathrm{median}}_{i \le j} \big( |X_i-X_j| \big).
\]
It is easy to show that the above test statistic {by \cite{Jeong/Son/Lee/Kim:2018} is also a pivotal quantity}.
However, it does not converge to the standard normal distribution. Thus,
\citeauthor{Jeong/Son/Lee/Kim:2018} suggested the following test statistic given by 
%----------------------------------------------------------------
\begin{equation} \label{EQ:TB}
T_{B} = \sqrt{\frac{6n}{\pi}} {\Phi^{-1}\Big( \frac{3}{4} \Big)}
  \frac{\displaystyle\mathop{\mathrm{median}}_{i \le j} \Big(\frac{X_i+X_j}{2}\Big) - \mu }%
{\displaystyle\mathop{\mathrm{median}}_{i \le j}\big(|X_i-X_j|\big)},
\end{equation}
%----------------------------------------------------------------
which converges to the standard normal distribution and is also pivotal; see Section 2.2 of \cite{Jeong/Son/Lee/Kim:2018}.

%===================================================================
\section{Empirical distributions} \label{section:03}
%-------------------------------------------------------------------
As aforementioned, the two robustified test statistics,  $T_A$ in (\ref{EQ:TA})
and $T_B$ in (\ref{EQ:TB}), converge to the standard normal distribution, assuming
that the sample size is large enough. Therefore, it is not appropriate to the
asymptotic properties of these statistics for making statistical inference, when the
sample sizes are small.
To deal with the difficulty for finding theoretical distributions of $T_A$
and $T_B$ for finite sample sizes, we carry out  extensive Monte Carlo simulations
to obtain the empirical distributions of $T_A$ and $T_B$ and calculate their
empirical quantiles, which can then be easily used for estimating critical values, confidence interval, $p$-value,
etc.

We used the R language~\citep{R:2018} to conduct simulation studies summarized as follows.
We generated one hundred million ($N=10^8$) samples from the standard normal distribution to obtain the empirical distributions of
$T_A$ and $T_B$ with the sample size $n$ ranging from $4$ to $50$ with an increment by one.
To be more specific,  for each given sample size, we used the generated samples to
obtain the empirical distributions of $T_A$ or $T_B$ and calculate their
empirical quantiles of $p$ by inverting the empirical distribution.
These empirical quantiles are provided in Table~\ref{TBL:quantilesTA} for $T_A$ and
Table~\ref{TBL:quantilesTB} for $T_B$ at the following lower quantiles of
0.6, 0.65, 0.7, 0.75, 0.8, 0.85, 0.9, 0.95, 0.975, 0.98, 0.99, and 0.995.

It is worthwhile to discuss the accuracy of the empirical quantiles obtained above.
Let $F^{-1}_N(p)$ be the
empirical quantile of $p$ obtained from the $N$ replications and
$F^{-1}(p)$ be the true quantile of $p$. Then it is easily seen from
Corollary~21.5 of \cite{Vaart:1998} that the sequence $\sqrt{N}\big(
F_N^{-1} - F^{-1}(p) \big)$ is asymptotically normal with mean zero and
variance $p/(1-p)/f^2\big(F^{-1}(p) \big)$. Thus, the standard deviation
of the empirical quantile of $p$, is approximately proportional to
$\sqrt{p(1-p)/N}$ which has its maximum value at $p=0.5$.  {Consequently},
the empirical quantiles are computed with an approximate accuracy of
$0.5/\sqrt{N}$. With $N=10^8$, we have $0.5/\sqrt{N}=0.00005$ which
roughly indicates that the empirical quantiles are accurate
up to the fourth decimal point.

Given that the probability density functions of $T_A$ and $T_B$ are symmetric at zero,
we have $F(-x) = 1 - F(x)$ and $F^{-1}(1/2)=0$.
{Letting $q_p$ be the $p$th lower quantile so that $F(q_p)=p$,
we} have $q_{1-p} = -q_{p}$.
Thus, it is enough to find the $p$th quantile only when $p>1/2$.
Let $G(\cdot)$ be the cumulative distribution function of $|X|$. Then we have
\[
G(x) = P[ |X| \le x ] = P[ -x \le X \le x ] = F(x) - F(-x)
     = 2F(x) -1.
\]
Substituting $x=q_p$ into the above, we have $G(q_p) = 2p-1$.
Thus, we have
\[
q_p = G^{-1}(2p-1),
\]
which is more effective than using $q_p = F^{-1}(p)$ in obtaining empirical quantiles.
In what follows, we illustrate the use of the empirical quantiles through some illustrative
examples.

%============================
\section{Illustrative examples} \label{section:04}
%----------------------------
In this section, we illustrate the applications of the obtained empirical quantiles for
constructing the confidence interval in Section \ref{section:04:01} and conducting statistical hypothesis testing
in Section \ref{section:04:02}.

\subsection{Confidence interval} \label{section:04:01}
%----------------------------
Let $\alpha_1$ and $\alpha_2$ with $\alpha=\alpha_1+\alpha_2$.
Let $q_{\alpha_1}$ and $q_{\alpha_2}$ be the $1-\alpha_1$ and $\alpha_2$ upper quantiles of
the distribution of {the statistic} $T_A$, respectively. Then we have
\[
P( q_{\alpha_1} \le T_A \le q_{\alpha_2} ) = 1 - \alpha.
\]
Thus, solving the following for $\mu$
\[
q_{\alpha_1} \le T_A \le q_{\alpha_2},
\]
we can obtain a $100(1-\alpha)$\% confidence interval for $\mu$ as follows
\[
\Bigg[
\mathop{\mathrm{median}}_{1\le i\le n}X_i
- \frac{ q_{\alpha_2} \sqrt{{\pi}/{2}}}{\Phi^{-1}\Big(\frac{3}{4}\Big)\sqrt{n}}
        \big|X_i-\mathop{\mathrm{median}}_{1\le i\le n}X_i\big|, ~
%---
\mathop{\mathrm{median}}_{1\le i\le n}X_i
- \frac{ q_{\alpha_1} \sqrt{{\pi}/{2}}}{\Phi^{-1}\Big(\frac{3}{4}\Big)\sqrt{n}}
        \big|X_i-\mathop{\mathrm{median}}_{1\le i\le n}X_i\big|
\Bigg].
\]
If we consider {the equi-tailed} confidence interval ($\alpha_1=\alpha_2=\alpha/2$),
then we have $q_{\alpha_1}=-q_{\alpha/2}$ and  $q_{\alpha_2}=q_{\alpha/2}$
since the distribution of $T_A$ is symmetric.
The end points of the confidence interval are given by
\[
\mathop{\mathrm{median}}_{1\le i\le n}X_i
\pm q_{\alpha/2} \frac{\sqrt{{\pi}/{2}}}{\Phi^{-1}\Big(\frac{3}{4}\Big)\sqrt{n}}
    \big|X_i-\mathop{\mathrm{median}}_{1\le i\le n}X_i\big|.
\]

{In a similar way as done above, we can also} obtain a  $100(1-\alpha)$\% confidence interval for $\mu$ using the statistic $T_B$,
which is given by
\begin{align*}
&\Bigg[
\mathop{\mathrm{median}}_{i \le j} \biggl(\frac{X_i+X_j}{2} \biggr)
- \frac{ q_{\alpha_2} \sqrt{{\pi}/{6}}}{\Phi^{-1}\Big(\frac{3}{4}\Big)\sqrt{n}}
    \mathop{\mathrm{median}}_{i \le j}\big(|X_i-X_j|\big), ~  \\
%---
&\qquad\qquad\qquad\qquad\qquad
\mathop{\mathrm{median}}_{i \le j} \biggl(\frac{X_i+X_j}{2} \biggr)
- \frac{ q_{\alpha_1} \sqrt{{\pi}/{6}}}{\Phi^{-1}\Big(\frac{3}{4}\Big)\sqrt{n}}
    \mathop{\mathrm{median}}_{i \le j}\big(|X_i-X_j|\big)
\Bigg],
\end{align*}
where $q_{\alpha_1}$ and $q_{\alpha_2}$ be the $1-\alpha_1$ and $\alpha_2$ upper quantiles of
the distribution of $T_B$, respectively.
The  end points of {the equi-tailed} confidence interval are also easily obtained as
\[
\mathop{\mathrm{median}}_{i \le j} \big(\frac{X_i+X_j}{2} \big)
\pm \frac{ q_{\alpha/2} \sqrt{{\pi}/{6}}}{\Phi^{-1}\Big(\frac{3}{4}\Big)\sqrt{n}}
    \mathop{\mathrm{median}}_{i \le j}\big(|X_i-X_j|\big).
\]

%%---------------------------------------------------
\begin{figure}[t]
\includegraphics{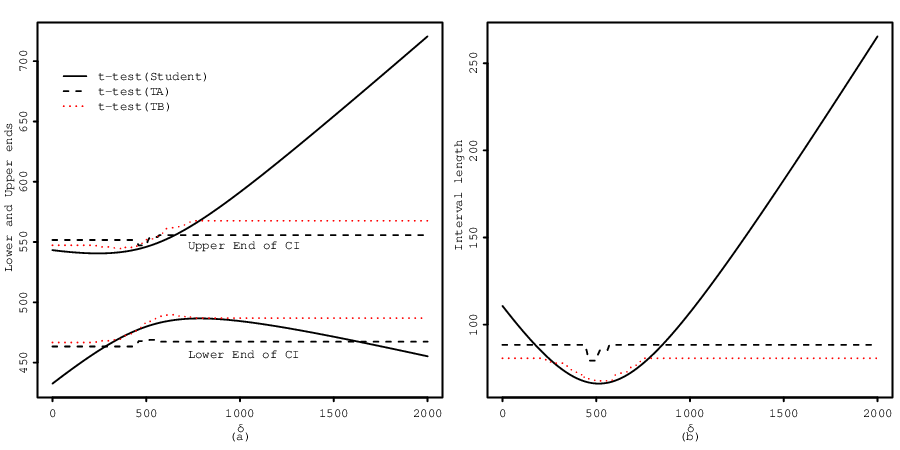}  %%-Figure
\caption{Confidence intervals and its corresponding interval length of
the three statistics.
(a) Confidence intervals. (b) Interval lengths.
}
\label{FIG:CI}
\end{figure}
%%---------------------------------------------------

As an illustration,
we consider the data set provided by Example 7.1-5 of \cite{Hogg/Tanis/Zimmerman:2015}.
In the example, the data about the amount of butterfat in pounds produced by a typical cow
are 481, 537, 513, 583, 453, 510, 570, 500, 457, 555,
       618, 327, 350, 643, 499, 421, 505, 637, 599, 392. Based on the normality assumption,
\citeauthor{Hogg/Tanis/Zimmerman:2015} obtained the confidence intervals based on the
$t$-test statistic in (\ref{t:test}), which is given by
\[
[472.80, ~542,20].
\]
To investigate the effect of data contamination,
we replaced the last observation (392)
with the value of $\delta$ ranging from $0$ to $2000$ in a grid-like fashion.
In Figure~\ref{FIG:CI} (a), we plotted the low and upper ends of the confidence interval
based on the Student $t$-test, $T_A$ and $T_B$ versus the value of $\delta$.
In Figure~\ref{FIG:CI} (b), we plotted the interval lengths of the  confidence intervals
under consideration. As shown in Figure~\ref{FIG:CI}, the confidence intervals based on the
conventional Student $t$-test statistic change dramatically while
the confidence intervals based on $T_A$ and  $T_B$ do not change much, justifying
the robustness of $T_A$ and  $T_B$.

%----------------------------
\subsection{Empirical powers} \label{section:04:02}
%----------------------------
Besides the confidence intervals, we can also easily
employ the robustified $t$-test statistics $T_A$ and $T_B$ to
conduct the hypothesis testing of the form
$H_0: \mu=0$ versus $H_a: \mu \neq 0$.
{In this subsection, we compare the empirical statistical powers of
these two statistics with the power using the $t$-test statistic in (\ref{t:test}).
Here, the statistical power of a hypothesis test is the probability
that the test correctly rejects the null hypothesis
when the alternative hypothesis is true.}

%%---------------------------------------------------
\begin{figure}[t]
\includegraphics{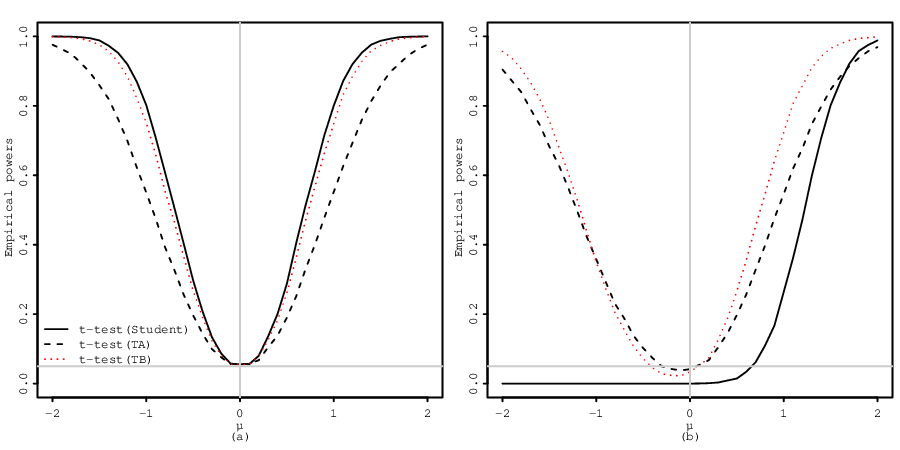}  %%-Figure
\caption{The empirical powers for $H_0:\mu=0$
versus  $H_a:\mu \neq 0$ with
sample size $n=10$.
(a) No contamination
and (b) Contamination ($x_1=10$).
}
\label{FIG:power11}
\end{figure}
%%---------------------------------------------------

To obtain the power curve of a hypothesis test,
we generated the first sample of size $n=10$ from
$N(\mu,1)$. The second sample of size $n=10$ is also generated from $N(\mu,1)$
but one observation in the sample is contaminated by assigning {the value of} $10$.
For a given value of $\mu$, we generated a sample and performed the hypothesis test.
We repeated this hypothesis test 10,000 times.
By calculating the number of {rejections of $H_0$}
divided by
the 10,000, we can obtain the empirical power at a given value of $\mu$.
The value of $\mu$ is changed from $-2$ to $2$ in a grid-like fashion.
These results are plotted in Figure~\ref{FIG:power11}.

As shown in Figure~\ref{FIG:power11} (a),
the empirical power using the $t$-test statistic in (\ref{t:test}) has
the highest in the absence of data contamination, as one expects.
It deserves pointing out that the power using $T_B$ is very close to that using the $t$-test statistic
and that the power using $T_A$ loses power noticeably. However, when there is contamination, the powers based on $T_A$ and $T_B$ are clearly higher than the one 
based on the $t$-test statistic as shown in Figure~\ref{FIG:power11} (b).

%% \clearpage
%%===============================
\section{Concluding remarks}  \label{section:05}
%%===============================
In this paper, we have carried out extensive Monte Carlo simulations
to obtain the empirical distributions of the two robustified test statistics,  $T_A$ in (\ref{EQ:TA})
and $T_B$ in (\ref{EQ:TB}) and calculate their
empirical quantiles, which can be easily used to improve statistical inference,
especially when the sample sizes are small or moderate. Thus, the obtained empirical quantiles could
further extend the use of these two tests in most practical applications.

It is worth pointing out that we only provided the empirical quantiles of
0.6, 0.65, 0.7, 0.75, 0.8, 0.85, 0.9, 0.95, 0.975, 0.98, 0.99, and 0.995
for $n=4,5,\ldots,50$. However, from a practical viewpoint,
these empirical quantiles are sufficient for most practical problems.
To make our results become more accessible to practitioners, 
we are currently updating the  \texttt{rt.test} R package \citep{Park/Wang:2018},
which provides all the detailed empirical quantiles, which should be
enough for calculating the confidence interval and the $p$-value.

%% %%===============================
%% \section*{Acknowledgment}
%% %%===============================
%% This work was supported by a 2-Year Research Grant of Pusan National University.
%% 
%% 

%%===============================================
\bibliographystyle{apalike}
\bibliography{REFmw10}
%%===============================================

%%==================================================================
\clearpage
 \vspace*{2in}
\section*{\centering\Huge Appendix: Tables for empirical quantiles}
%%---------------------------------------------------

\clearpage
%% \thispagestyle{empty}
%%---------------------------------------------------
%%\setlength\headheight{-75pt}
%% \setlength\topmargin{-20pt} \setlength\oddsidemargin{-10pt} \setlength\evensidemargin{-10pt}
\renewcommand{\baselinestretch}{1.00}
\begin{table}[t!]
\begin{center}
\caption{Empirical quantiles of $T_A$ statistic with sample size $n$,
where $n=4,5,\ldots,50$.}\label{TBL:quantilesTA}

\smallskip
\begin{footnotesize}
\begin{tabular}{cl*{12}{r}} %\\[-0.5ex]
\hline %% \\[-2ex]
   && \multicolumn{11}{c}{$p$}  & \\
$n$&& 0.60 & 0.65 & 0.70 & 0.75 & 0.80 & 0.85 & 0.90 & 0.95 & 0.975 & 0.98 & 0.99 & 0.995 \\
\cline{1-1}\cline{3-14}
4 && 0.298 &0.462 &0.649 &0.856 &1.113 &1.480 &2.054 &3.342 &5.096 &5.826 &8.265 &11.494 \\
5 && 0.305 &0.466 &0.646 &0.852 &1.128 &1.471 &2.046 &3.305 &5.073 &5.660 &8.172 &12.239 \\
6 && 0.276 &0.422 &0.574 &0.762 &0.981 &1.272 &1.691 &2.510 &3.491 &3.835 &5.224 & 6.816 \\
7 && 0.282 &0.435 &0.599 &0.788 &1.020 &1.291 &1.748 &2.546 &3.556 &3.946 &5.198 & 6.860 \\
8 && 0.269 &0.407 &0.563 &0.727 &0.940 &1.209 &1.563 &2.218 &2.924 &3.161 &4.171 & 5.320 \\
9 && 0.279 &0.435 &0.587 &0.758 &0.968 &1.229 &1.611 &2.282 &3.030 &3.263 &4.205 & 5.406 \\
10&& 0.266 &0.402 &0.553 &0.710 &0.906 &1.145 &1.490 &2.056 &2.627 &2.848 &3.565 & 4.369 \\
11&& 0.270 &0.413 &0.567 &0.738 &0.936 &1.182 &1.520 &2.106 &2.729 &2.942 &3.652 & 4.449 \\
12&& 0.259 &0.396 &0.543 &0.706 &0.894 &1.125 &1.440 &1.971 &2.520 &2.704 &3.301 & 3.952 \\
13&& 0.267 &0.408 &0.560 &0.727 &0.919 &1.155 &1.475 &2.013 &2.569 &2.755 &3.360 & 4.022 \\
14&& 0.258 &0.394 &0.540 &0.700 &0.885 &1.109 &1.411 &1.912 &2.418 &2.584 &3.118 & 3.687 \\
15&& 0.265 &0.405 &0.554 &0.719 &0.907 &1.137 &1.444 &1.952 &2.464 &2.632 &3.173 & 3.749 \\
16&& 0.257 &0.393 &0.537 &0.696 &0.878 &1.098 &1.392 &1.871 &2.346 &2.500 &2.991 & 3.505 \\
17&& 0.264 &0.402 &0.550 &0.713 &0.899 &1.123 &1.421 &1.908 &2.388 &2.545 &3.041 & 3.561 \\
18&& 0.257 &0.392 &0.535 &0.693 &0.873 &1.090 &1.377 &1.841 &2.294 &2.440 &2.899 & 3.373 \\
19&& 0.263 &0.401 &0.547 &0.709 &0.892 &1.113 &1.405 &1.875 &2.333 &2.480 &2.944 & 3.423 \\
20&& 0.256 &0.391 &0.534 &0.691 &0.869 &1.083 &1.366 &1.818 &2.254 &2.393 &2.828 & 3.273 \\
21&& 0.262 &0.399 &0.545 &0.705 &0.887 &1.105 &1.392 &1.849 &2.290 &2.431 &2.870 & 3.319 \\
22&& 0.256 &0.390 &0.533 &0.689 &0.866 &1.079 &1.357 &1.799 &2.222 &2.356 &2.773 & 3.194 \\
23&& 0.261 &0.398 &0.543 &0.702 &0.883 &1.098 &1.381 &1.829 &2.256 &2.392 &2.812 & 3.238 \\
24&& 0.256 &0.390 &0.532 &0.688 &0.864 &1.074 &1.349 &1.784 &2.196 &2.326 &2.727 & 3.132 \\
25&& 0.260 &0.397 &0.542 &0.700 &0.879 &1.093 &1.372 &1.812 &2.228 &2.360 &2.765 & 3.172 \\
26&& 0.255 &0.389 &0.531 &0.686 &0.862 &1.071 &1.344 &1.772 &2.175 &2.301 &2.691 & 3.081 \\
27&& 0.260 &0.396 &0.540 &0.698 &0.876 &1.088 &1.364 &1.798 &2.205 &2.333 &2.727 & 3.119 \\
28&& 0.255 &0.389 &0.531 &0.685 &0.860 &1.068 &1.339 &1.762 &2.157 &2.281 &2.661 & 3.037 \\
29&& 0.259 &0.395 &0.539 &0.696 &0.874 &1.084 &1.358 &1.786 &2.186 &2.311 &2.694 & 3.074 \\
30&& 0.255 &0.388 &0.530 &0.685 &0.859 &1.066 &1.334 &1.753 &2.142 &2.264 &2.635 & 3.001 \\
31&& 0.259 &0.394 &0.538 &0.695 &0.871 &1.081 &1.353 &1.776 &2.169 &2.292 &2.666 & 3.035 \\
32&& 0.255 &0.388 &0.530 &0.684 &0.858 &1.064 &1.331 &1.745 &2.129 &2.249 &2.613 & 2.970 \\
33&& 0.258 &0.394 &0.537 &0.693 &0.869 &1.078 &1.348 &1.767 &2.155 &2.276 &2.642 & 3.002 \\
34&& 0.255 &0.388 &0.529 &0.683 &0.857 &1.062 &1.327 &1.739 &2.118 &2.235 &2.593 & 2.943 \\
35&& 0.258 &0.393 &0.536 &0.692 &0.868 &1.076 &1.344 &1.760 &2.142 &2.261 &2.621 & 2.973 \\
36&& 0.255 &0.388 &0.529 &0.683 &0.856 &1.060 &1.324 &1.733 &2.108 &2.225 &2.576 & 2.919 \\
37&& 0.258 &0.393 &0.536 &0.691 &0.866 &1.074 &1.341 &1.753 &2.131 &2.249 &2.603 & 2.948 \\
38&& 0.254 &0.388 &0.529 &0.682 &0.855 &1.059 &1.322 &1.728 &2.099 &2.214 &2.562 & 2.898 \\
39&& 0.258 &0.393 &0.535 &0.690 &0.865 &1.071 &1.337 &1.747 &2.122 &2.238 &2.588 & 2.927 \\
40&& 0.254 &0.388 &0.529 &0.682 &0.854 &1.058 &1.320 &1.723 &2.091 &2.205 &2.548 & 2.880 \\
41&& 0.257 &0.392 &0.535 &0.690 &0.864 &1.070 &1.334 &1.741 &2.113 &2.228 &2.573 & 2.908 \\
42&& 0.254 &0.388 &0.528 &0.681 &0.853 &1.057 &1.318 &1.719 &2.085 &2.198 &2.537 & 2.864 \\
43&& 0.257 &0.392 &0.534 &0.689 &0.863 &1.068 &1.332 &1.736 &2.105 &2.219 &2.560 & 2.889 \\
44&& 0.254 &0.387 &0.528 &0.681 &0.853 &1.056 &1.316 &1.716 &2.079 &2.190 &2.526 & 2.849 \\
45&& 0.257 &0.392 &0.534 &0.688 &0.862 &1.067 &1.329 &1.732 &2.098 &2.211 &2.549 & 2.874 \\
46&& 0.254 &0.387 &0.528 &0.681 &0.852 &1.055 &1.314 &1.712 &2.073 &2.184 &2.516 & 2.835 \\
47&& 0.257 &0.391 &0.533 &0.688 &0.861 &1.065 &1.327 &1.728 &2.092 &2.204 &2.539 & 2.861 \\
48&& 0.254 &0.387 &0.528 &0.680 &0.852 &1.054 &1.313 &1.709 &2.068 &2.178 &2.508 & 2.824 \\
49&& 0.257 &0.391 &0.533 &0.687 &0.860 &1.064 &1.325 &1.725 &2.086 &2.197 &2.529 & 2.847 \\
50&& 0.254 &0.387 &0.528 &0.680 &0.851 &1.053 &1.312 &1.706 &2.063 &2.172 &2.499 & 2.813 \\
%----------------------------
\hline
\end{tabular}
\end{footnotesize}
\end{center}
\end{table}
%-------------------------------------------------

\clearpage
%% \thispagestyle{empty}
%%---------------------------------------------------
%%\setlength\headheight{-75pt}
%% \setlength\topmargin{-20pt} \setlength\oddsidemargin{-10pt} \setlength\evensidemargin{-10pt}
\renewcommand{\baselinestretch}{1.00}
\begin{table}[t!]
\begin{center}
\caption{Empirical quantiles of $T_B$ statistic with sample size $n$,
where $n=4,5,\ldots,50$.}\label{TBL:quantilesTB}

\bigskip
\begin{footnotesize}
\begin{tabular}{cl*{12}{r}} %\\[-0.5ex]
\hline %% \\[-2ex]
   && \multicolumn{11}{c}{$p$}  & \\
$n$&& 0.60 & 0.65 & 0.70 & 0.75 & 0.80 & 0.85 & 0.90 & 0.95 & 0.975 & 0.98 & 0.99 & 0.995 \\
\cline{1-1}\cline{3-14}
4 && 0.215 &0.330 &0.455 &0.597 &0.765 &0.979 &1.286 &1.853 &2.511 &2.748 &3.588 &4.619 \\
5 && 0.232 &0.356 &0.491 &0.642 &0.822 &1.050 &1.376 &1.979 &2.680 &2.935 &3.841 &4.962 \\
6 && 0.233 &0.356 &0.489 &0.637 &0.810 &1.025 &1.322 &1.843 &2.406 &2.601 &3.263 &4.028 \\
7 && 0.240 &0.367 &0.503 &0.654 &0.828 &1.042 &1.333 &1.826 &2.339 &2.512 &3.081 &3.710 \\
8 && 0.241 &0.368 &0.504 &0.654 &0.827 &1.037 &1.321 &1.796 &2.279 &2.440 &2.964 &3.537 \\
9 && 0.242 &0.369 &0.505 &0.655 &0.826 &1.034 &1.311 &1.765 &2.218 &2.366 &2.840 &3.345 \\
10&& 0.244 &0.372 &0.508 &0.658 &0.830 &1.036 &1.309 &1.752 &2.186 &2.327 &2.771 &3.234 \\
11&& 0.245 &0.373 &0.510 &0.660 &0.831 &1.036 &1.307 &1.742 &2.164 &2.299 &2.724 &3.162 \\
12&& 0.245 &0.374 &0.511 &0.661 &0.831 &1.035 &1.303 &1.730 &2.138 &2.269 &2.676 &3.089 \\
13&& 0.246 &0.375 &0.512 &0.662 &0.832 &1.035 &1.301 &1.721 &2.120 &2.247 &2.639 &3.035 \\
14&& 0.247 &0.376 &0.514 &0.664 &0.833 &1.036 &1.300 &1.716 &2.108 &2.232 &2.613 &2.996 \\
15&& 0.247 &0.377 &0.514 &0.664 &0.834 &1.036 &1.298 &1.710 &2.096 &2.217 &2.588 &2.959 \\
16&& 0.248 &0.377 &0.515 &0.665 &0.834 &1.035 &1.296 &1.704 &2.084 &2.203 &2.566 &2.926 \\
17&& 0.248 &0.378 &0.515 &0.665 &0.834 &1.035 &1.295 &1.700 &2.075 &2.192 &2.549 &2.901 \\
18&& 0.248 &0.378 &0.516 &0.666 &0.835 &1.036 &1.295 &1.697 &2.068 &2.184 &2.535 &2.880 \\
19&& 0.248 &0.379 &0.516 &0.666 &0.835 &1.035 &1.294 &1.694 &2.062 &2.176 &2.522 &2.860 \\
20&& 0.249 &0.379 &0.517 &0.666 &0.835 &1.035 &1.293 &1.690 &2.055 &2.168 &2.509 &2.841 \\
21&& 0.249 &0.379 &0.517 &0.667 &0.836 &1.035 &1.292 &1.688 &2.050 &2.162 &2.499 &2.827 \\
22&& 0.249 &0.380 &0.518 &0.667 &0.836 &1.035 &1.291 &1.686 &2.045 &2.156 &2.490 &2.813 \\
23&& 0.249 &0.380 &0.518 &0.668 &0.836 &1.035 &1.291 &1.684 &2.041 &2.151 &2.482 &2.801 \\
24&& 0.250 &0.380 &0.518 &0.668 &0.836 &1.035 &1.291 &1.682 &2.037 &2.147 &2.475 &2.790 \\
25&& 0.250 &0.380 &0.518 &0.668 &0.837 &1.035 &1.290 &1.680 &2.033 &2.142 &2.467 &2.780 \\
26&& 0.250 &0.380 &0.519 &0.669 &0.837 &1.036 &1.290 &1.679 &2.031 &2.138 &2.461 &2.771 \\
27&& 0.250 &0.381 &0.519 &0.669 &0.837 &1.035 &1.289 &1.677 &2.027 &2.134 &2.455 &2.762 \\
28&& 0.250 &0.381 &0.519 &0.669 &0.837 &1.035 &1.289 &1.675 &2.024 &2.131 &2.450 &2.754 \\
29&& 0.250 &0.381 &0.519 &0.669 &0.837 &1.035 &1.289 &1.674 &2.022 &2.128 &2.445 &2.747 \\
30&& 0.250 &0.381 &0.519 &0.669 &0.838 &1.036 &1.289 &1.674 &2.020 &2.126 &2.441 &2.741 \\
31&& 0.250 &0.381 &0.519 &0.669 &0.838 &1.036 &1.288 &1.672 &2.017 &2.123 &2.437 &2.735 \\
32&& 0.250 &0.381 &0.520 &0.669 &0.838 &1.035 &1.288 &1.671 &2.016 &2.121 &2.433 &2.730 \\
33&& 0.251 &0.381 &0.520 &0.670 &0.838 &1.036 &1.288 &1.670 &2.014 &2.118 &2.429 &2.724 \\
34&& 0.251 &0.382 &0.520 &0.670 &0.838 &1.036 &1.288 &1.670 &2.012 &2.117 &2.426 &2.719 \\
35&& 0.251 &0.382 &0.520 &0.670 &0.838 &1.036 &1.287 &1.669 &2.011 &2.115 &2.423 &2.715 \\
36&& 0.251 &0.382 &0.520 &0.670 &0.838 &1.035 &1.287 &1.668 &2.008 &2.112 &2.419 &2.709 \\
37&& 0.251 &0.382 &0.520 &0.670 &0.838 &1.036 &1.287 &1.667 &2.007 &2.111 &2.417 &2.706 \\
38&& 0.251 &0.382 &0.520 &0.670 &0.838 &1.036 &1.287 &1.667 &2.006 &2.109 &2.414 &2.702 \\
39&& 0.251 &0.382 &0.520 &0.670 &0.838 &1.036 &1.287 &1.666 &2.005 &2.107 &2.412 &2.699 \\
40&& 0.251 &0.382 &0.521 &0.671 &0.838 &1.036 &1.287 &1.666 &2.004 &2.106 &2.410 &2.696 \\
41&& 0.251 &0.382 &0.521 &0.671 &0.838 &1.036 &1.286 &1.665 &2.002 &2.105 &2.407 &2.692 \\
42&& 0.251 &0.382 &0.521 &0.671 &0.839 &1.036 &1.286 &1.665 &2.001 &2.103 &2.405 &2.689 \\
43&& 0.251 &0.382 &0.521 &0.671 &0.839 &1.036 &1.286 &1.664 &2.000 &2.102 &2.403 &2.686 \\
44&& 0.251 &0.382 &0.521 &0.671 &0.839 &1.036 &1.286 &1.663 &1.999 &2.101 &2.401 &2.683 \\
45&& 0.251 &0.382 &0.521 &0.671 &0.839 &1.036 &1.286 &1.663 &1.998 &2.100 &2.399 &2.681 \\
46&& 0.251 &0.382 &0.521 &0.671 &0.839 &1.036 &1.285 &1.662 &1.997 &2.098 &2.398 &2.679 \\
47&& 0.251 &0.382 &0.521 &0.671 &0.839 &1.036 &1.285 &1.662 &1.996 &2.097 &2.395 &2.676 \\
48&& 0.251 &0.383 &0.521 &0.671 &0.839 &1.036 &1.286 &1.662 &1.995 &2.096 &2.394 &2.674 \\
49&& 0.252 &0.383 &0.521 &0.671 &0.839 &1.036 &1.285 &1.662 &1.995 &2.096 &2.393 &2.672 \\
50&& 0.252 &0.383 &0.521 &0.671 &0.839 &1.036 &1.285 &1.661 &1.994 &2.095 &2.392 &2.669 \\
%----------------------------
\hline
\end{tabular}
\end{footnotesize}
\end{center}
\end{table}
%-------------------------------------------------

%%===============================================
\end{document}